\documentclass{jetpl}
\twocolumn
\lat
\usepackage{cite}
\usepackage{graphicx}
\usepackage{url}
\renewcommand{\D}{\mathrm d}
\newcommand{\E}{\mathrm e}
\newcommand{\I}{\mathrm i}

\title{Direct numerical experiment on measuring of dispersion relation for gravity waves in the presence of condensate}

\rtitle{Direct measurement of dispersion relation\ldots}

\sodtitle{Direct numerical experiment on measuring of dispersion relation for gravity waves in the presence of condensate}

\author{A.\,O.\,Korotkevich$^{+*}$\/\thanks{e-mail: alexkor@math.unm.edu}}

\rauthor{A.\,O.\,Korotkevich}

\sodauthor{Korotkevich}

\address{$^+$Department of Mathematics and Statistics, University of New Mexico, MSC01 1115, 1 University of New Mexico, Albuquerque, NM 87131-0001, USA\\~\\
$^*$L.\,D.~Landau Institute for Theoretical Physics RAS, 2 Kosygin Str., Moscow, 119334, Russian Federation}

\dates{2 December 2012}{*}

\abstract{During previous numerical experiments on isotropic turbulence of surface gravity waves
we observed formation of the long wave background (condensate). It was shown (Korotkevich, Phys.\,Rev.\,Lett.~{\bf 101}(7), 074504 (2008)),
that presence of the condensate
changes a spectrum of direct cascade, corresponding to the flux of energy to the small scales
from pumping region (large scales). Recent experiments show that the inverse cascade spectrum
is also affected by the condensate. In this case mechanism proposed as a cause for the change of direct cascade
spectrum cannot work. But inverse cascade is directly influenced by the linear dispersion relation for waves,
as a result direct measurement of the dispersion relation in the presence of condensate is necessary.
We performed the measurement of this dispersion relation from the direct numerical experiment. The results demonstrate
that in the region of inverse cascade influence of the condensate cannot be neglected.}

\PACS{47.27.E-, 47.27.Gs, 47.35.Bb}
\begin{document}

\maketitle

Theory of wave or weak turbulence~\cite{ZLF1992} applied to gravity waves on the surface of the fluid
is the base for all current wave forecasting models, which are
crucial for ocean cargo communications and oil and gas sea platforms operations. This is why it's verification
is important and urgent problem. Numerous attempts to get a spectrum of the direct cascade of energy
from ocean and sea observations give results which confirm the wave turbulence theory~\cite{Donelan1985, Hwang2000}.
At the same time all these experiments were working with wind generated waves, which means broad spectrum
pumping. As a result it is hard to understand where we have so called ``inertial interval'' (region of scale
where we have only nonlinear interaction of waves, while pumping and damping influences are negligible).
Narrow frequency pumping can be realized in experimental wave tanks and flumes. But this state of the art
experiments were producing very strange and contradictory results~\cite{Lukaschuk2007, FFL2007}, like dependence
of the spectral slope on the amplitude of pumping. The direct numerical simulation of the primordial
dynamical equations looks like a natural remedy for this problem. It provides us with all possible
information about the system, but for the cost of enormous computational complexity.

One of the first attempts was pioneering work~\cite{Onorato2002}, which soon was followed
by~\cite{DKZ2003grav,DKZ2003gravrus,DKZ2004}
During the last decade,
the author together with colleagues were able to find answers at least to some of the open questions
using direct numerical simulation of gravity waves.
It was shown, that on a discrete grid of wavenumbers (common situation for both pseudo spectral numerical
codes in a periodic domain and experimental wave tanks which are usually rectangular basins of finite size)
the mesoscopic turbulence can take place~\cite{ZKPD2005, ZKPD2005rus}.
In the recent works~\cite{Korotkevich2008PRL, Korotkevich2012MCS} the author demonstrated that formation of
the inverse cascade, corresponding to the flux of wave action (analog of number of waves), inevitably
leads to the formation of the strong long wave background, which we call condensate (due to similarity
with Bose-Einstein condensation in condensed matter physics). It was shown that presence of the condensate
changes the slope of the direct cascade spectrum. In these recent experiments, in spite of the short
inertial interval for the inverse cascade, the slope significantly different from the predicted by
the wave turbulence theory was observed. Recent reports~\cite{Korotkevich2012Loughborough} with long enough inertial interval show that
the slope is indeed differs from the theoretically predicted one.

In the present Letter we report results of direct measurement of the dispersion relation of the surface gravity waves
in the presence of condensate. As it will be demonstrated later, the slope of the spectrum of the inverse cascade
directly depends on the power of the dispersion relation. Although we were not able to determine the change of
this power, we demonstrate that in the region of the inverse cascade the dispersion relation is strongly
affected by the nonlinear interaction with the condensate. This distortion cannot be neglected and might be considered as
a possible cause of the change of the slope of the inverse cascade spectrum.

As in many previous works~\cite{PZ1996, DKZ2003grav, DKZ2004, Korotkevich2008PRL, Korotkevich2012MCS} we shall consider
isotropic turbulence (no direct dependence on angle) of gravity waves on the 2D surface of the 3D fluid.
The isotropic turbulence is a wonderful sandbox for the turbulence simulation. In this case we can use angle averaging
of the resulting spectra in order to decrease it natural peakedness of harmonics, instead of averaging over
different realizations which is unrealistic in this case, due to a long time of even one simulation.

Here and further we shall follow notations from~\cite{Korotkevich2008PRL}.
We consider a potential flow of ideal incompressible fluid. System is described in terms of weakly nonlinear
equations~\cite{ZLF1992} for surface elevation $\eta(\vec r, t)$ and velocity potential at the surface $\psi(\vec r,t)$ ($\vec r = \overrightarrow{(x,y)}$)
\begin{align}
\dot \eta =& \hat k  \psi - (\nabla (\eta \nabla \psi)) - \hat k  [\eta \hat k  \psi]
+ \hat k (\eta \hat k  [\eta \hat k  \psi]) +\nonumber\\
+&\frac{1}{2} \Delta [\eta^2 \hat k \psi] + 
\frac{1}{2} \hat k [\eta^2 \Delta\psi] + \widehat F^{-1} [\gamma_k \eta_k],\nonumber\\
\dot \psi =& - g\eta - \frac{1}{2}\left[ (\nabla \psi)^2 - (\hat k \psi)^2 \right]
- [\hat k  \psi] \hat k  [\eta \hat k  \psi] -\label{eta_psi_system}\\
-& [\eta \hat k  \psi]\Delta\psi + \widehat F^{-1} [\gamma_k \psi_k] + P_{\vec r}.\nonumber
\end{align}
Here dot means time-derivative, $\Delta$ --- Laplace operator, $\hat k$ is a linear integral operator
$\left(\hat k =\sqrt{-\Delta}\right)$, $\widehat F^{-1}$ is an inverse Fourier transform, $\gamma_k$ is a
dissipation rate (according to recent work~\cite{DDZ2008} it has to be included in both equations), which
corresponds to viscosity
on small scales and, if needed, "artificial" damping on large scales.
$P_{\vec r}$ is the driving term which simulates pumping on large scales (for example, due to wind).
In the $k$-space supports of $\gamma_{k}$ and $P_{\vec k}$ are separated by the inertial
interval, where the Kolmogorov-type solution can be recognized.
These equations were derived as a results of Hamiltonian expansion in terms of $\hat k \eta$.
From physical point of view
$\hat k$-operator is close to derivative, so we expand in powers of slope of the surface. In most of
experimental observations average slope of the open sea surface $\mu$ is of the order of $0.1$, so
such expansion is very reasonable. Additional details can be found in~\cite{Zakharov1999, DKZ2003cap, DKZ2003caprus, DKZ2003grav}.

In the case of statistical description of the wave field, Hasselmann kinetic equation~\cite{Hasselmann1962} for
the distribution of the wave action $n(k,t)\simeq\langle|a_{\vec{k}}(t)|^2\rangle$ is used. Here
\begin{equation}
a_{\vec k} = \sqrt \frac{\omega_k}{2k} \eta_{\vec k} + \I \sqrt \frac{k}{2\omega_k} \psi_{\vec k},
\end{equation}
are complex normal variables. For gravity waves $\omega_k = \sqrt{gk}$. In this variables, if we have
a linear wave with wavenumber $\vec k$, it will correspond to the only excited harmonics $a_{\vec k}$.
In other words, representation in terms of these normal variables means representation in terms of elementary
excitations in the system (linear waves). In reality we should use pair correlator for variables after canonical
transformation which eliminates nonresonant terms in the Hamiltonian~\cite{ZLF1992, Zakharov1999},
but in the case of gravity waves of average steepness $\sqrt{\langle|\nabla\eta|^2\rangle}\simeq 0.1$ their
relative difference is of the order of few percents. Thus we neglect this difference and will be working
with correlation function given above.

\section{Numerical simulation.}
We simulated primordial dynamical equations (\ref{eta_psi_system}) in a periodic spatial domain $2\pi\times 2\pi$.
Main part of the simulations was performed on a grid consisting of $1024\times 1024$ knots. Also we performed long
time simulation on the grid $256\times 256$.
The used numerical code~\cite{KorotkevichPhD} was verified in~\cite{DKZ2003cap, DKZ2003grav, DKZ2004, ZKPD2005, ZKPR2007, KPRZ2008, Korotkevich2008PRL, Korotkevich2012MCS}.
Gravity acceleration was $g=1$. Pseudo-viscous damping coefficient had the following form
\begin{equation}
\gamma_{k} = 
\begin{cases}
0, k \le k_d,\\
- \gamma_0 (k - k_d)^2, k > k_d,
\end{cases}
\end{equation}
where $k_d = 256$ and $\gamma_{0,1024} = 2.7\times10^{4}$ for the grid $1024\times 1024$ and $k_d = 64$
and $\gamma_{0,256} = 2.4\times10^{2}$ for the smaller grid $256\times 256$. Pumping was an isotropic driving
force narrow in wavenumbers space with random phase:
\begin{equation}
P_{\vec k} = f_k \E^{\I R_{\vec k} (t)}, f_k =
\begin{cases}
4 F_0 \frac{(k-k_{p1})(k_{p2}-k)}{(k_{p2} - k_{p1})^2},\\
0 - \mathrm{if}\; k < k_{p1}\;\mathrm{or}\; k > k_{p2};\\
\end{cases}
\end{equation}
here $k_{p1}=28,\; k_{p2}=32$ and $F_0 = 1.5\times 10^{-5}$; $R_{\vec k}(t)$ was uniformly distributed random
number in the interval $(0,2\pi]$ for each $\vec k$ and $t$. Initial condition was low amplitude noise in all
harmonics. Time steps were $\Delta t_{1024} = 6.7\times 10^{-4}$ and $\Delta t_{256} = 5.0\times 10^{-3}$.
We used Fourier series in the following form:
\begin{align*}
\eta_{\vec k} =& \widehat F[\eta_{\vec r}] = \frac{1}{(2\pi)^2}\int\limits_{0}^{2\pi}\int\limits_{0}^{2\pi}
\eta_{\vec r} \E^{\I \vec k\vec r}\D^2 r,\\
\eta_{\vec r} =& \widehat F^{-1}[\eta_{\vec k}] = \sum\limits_{-N_x/2}^{N_x/2}\sum\limits_{-N_y/2}^{N_y/2}
\eta_{\vec k} \E^{-\I \vec k\vec r},
\end{align*}
here $N_x$, $N_y$ --- are number of Fourier modes in $x$ and $y$ directions.

\begin{figure}[ht!]
\centering
\includegraphics[width=3.5in]{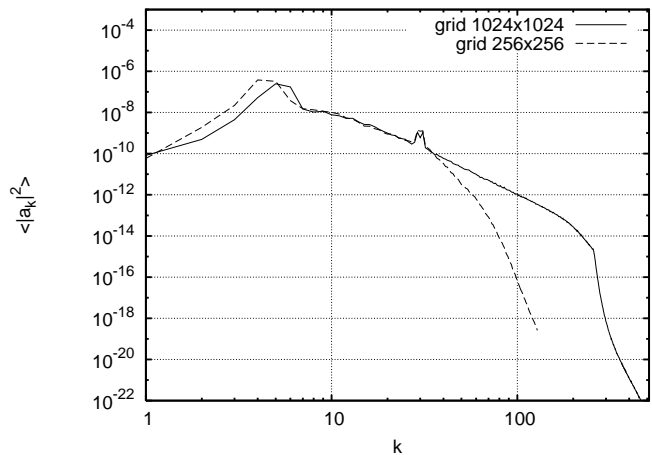}
\caption{\label{Spectra_all}Fig.~\ref{Spectra_all}. Spectra $\langle |a_k|^2\rangle$. With condensate on the $1024\times 1024$ grid (solid); on the $256\times 256$ grid with more developed condensate (long dashes).}
\end{figure}

As a results of simulation we observed~\cite{Korotkevich2008PRL, Korotkevich2012MCS} formation of both direct and inverse cascades (Fig.~\ref{Spectra_all}, solid line). Average steepness was equal to $\sqrt{\langle|\nabla \eta|^2\rangle} = 0.14$.
What is important, development of inverse cascade spectrum was arrested by
discreteness of wavenumbers grid in agreement with~\cite{DKZ2003cap, ZKPD2005, Nazarenko2006, LNP2006}. Then large scale
condensate started to form. The mechanism of condensate formation is the following. We have a flux of wave action (number
of waves) from the pumping to the large scale region. This flux is due to nonlinear resonant interaction of waves.
In~\cite{ZKPD2005} it was shown, that on a discrete grid of wavenumbers, which is typical for both finite experimental
wave tanks and computer simulations, resonance conditions are never fulfilled exactly. What makes it possible for them
to exist is the finite width of the resonant curve due to nonlinear frequency shift. As a result, this thick resonant
curve covers some knots of the wavenumbers grid~\cite{DKZ2003cap}. The nonlinear frequency shift is proportional to
the matrix element (coupling coefficient) of interaction of waves, which is homogeneous function of the order 3 (with
change of $k$ it behaves as $k^3$). Which means that we have good coverage of the grid knots in high wavenumbers,
characteristic for the direct cascade, and worse and worse situation as we move toward the origin of the $k$-plane.
So at some stage the resonant interactions turn off due to discreteness of the wavenumbers grid and flux cannot propagate
further. At the same time, new wave are still brought to this threshold scale from the pumping region. As a result we
have an accumulation of waves at some large scale, i.\,e.~condensation.

As one can see, value of wave action $|a_k|^2$ at the condensate region is more than
order of magnitude larger than for closest harmonic of inverse cascade. Dynamics of large scales
became extremely slow after this point. We managed to achieve downshift of condensate peak for one step of
wavenumbers grid during long time simulation on a small grid $256\times 256$
(Fig.~\ref{Spectra_all}, line with long dashes). As one can see we observed elongation of inverse cascade interval
without noticeable change of the slope. Unfortunately, inertial interval for inverse cascade is too short to
exclude possible influence of pumping and condensate, we can roughly estimate its slope as $k^{-3.5}$~\cite{Korotkevich2008PRL},
which is slightly less than the prediction of the theory of weak turbulence $k^{-23/6}\sim k^{-3.83}$.
Resent findings~\cite{Korotkevich2012Loughborough}
with significantly longer inertial interval for the inverse cascade, support this result and propose the slope $k^{-3.15}$.

Let us discuss possible reason for this deviation from the theory of weak turbulence. The direct cascade of energy and inverse
cascade of wave action correspond two Kolmogorov-type solutions of the Hasselmann kinetic equation for
waves~\cite{Hasselmann1962}. These solutions were derived by Zakharov in the middle of sixties under a few reasonable
assumptions: media is isotropic with respect to rotations, dispersion relation is a power-like function $\omega_k ~ k^{\alpha}$,
matrix element of nonlinear interaction (nonlinear coupling coefficient) for waves is a homogeneous function
$T(\varepsilon\vec k_1, \varepsilon\vec k_2, \varepsilon\vec k_3, \varepsilon\vec k_4) = \varepsilon^{\beta} T(\vec k_1, \vec k_2, \vec k_3, \vec k_4)$. Under these assumptions Zakharov~\cite{ZF1967,ZakharovPhD,ZZ1982,ZLF1992} obtained Kolmogorov-like solutions
corresponding to fluxes of two integrals of motion (energy and wave action or number of waves):
\begin{equation}
n_k^{(1)} = C_1 P^{1/3} k^{-\frac{2\beta}{3} - d},\;\;
n_k^{(2)} = C_2 Q^{1/3} k^{-\frac{2\beta - \alpha}{3} - d}.
\end{equation}
Here $d$ is a spatial dimension ($d=2$ in our case).
In the case of gravity waves on a deep water $\omega=\sqrt{gk}$ ($\alpha=1/2$) and, apparently, $\beta=3$.
As a result one can get:
\begin{equation}
\label{weak_turbulent_exponents}
n_k^{(1)} = C_1 P^{1/3} k^{-4},\;
n_k^{(2)} = C_2 Q^{1/3} k^{-23/6}.
\end{equation}
The first spectrum $n_k^{(1)}$ corresponds to the direct cascade. In this work we are more interested in
the second spectrum $n_k^{(2)}$ describes to inverse cascade, corresponding to flux of number of waves (or wave action)
from small scales (pumping) to larger scales. As one can see, inverse cascade spectrum formula directly
depends on $\alpha$, the power of the dispersion relation. So it would be helpful for understanding of the situation
to measure directly the dispersion relation of waves. How one can do this measurement?

In the case of linear plane wave, for harmonic $a_{\vec k}$ will have just rotation of phase with constant amplitude
$a_{\vec k}(t) = A_{\vec k}\exp(-\I\omega_k t)$,
with circular frequency corresponding the linear dispersion relation $\omega(\vec k) = \sqrt{gk}$.
So if one would write down the $a_{\vec k}(t)$ as a function on time and then calculate Fourier transform from
time domain to the frequency domain, the result of such transformation will be $\delta$-function in $\omega$-space.

\begin{figure}[ht!]
\centering
\includegraphics[width=3.4in]{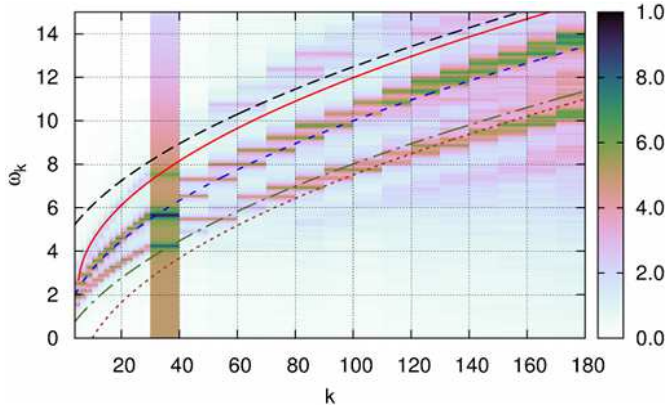}
\caption{\label{a_k_omega_surf}Fig.~\ref{a_k_omega_surf}. (Color online) Surface $|a_{\vec k}(\omega)|$. Normed to the maximum
for every given value of $k$. Bright verical line -- pumping region.
Sidebands, corresponding to the interaction with condensate, are clearly visible.
Sidebands estimations: long dashes (black) -- $\omega_k^{up+} = \omega_{k+k_c} + \omega_{k_c}$; solid line (red) -- $\omega_k^{up-} = \omega_{k-k_c} + \omega_{k_c}$; dashed-dotted line (olive) -- $\omega_k^{low+} = \omega_{k+k_c} + \omega_{k_c}$; dotted line (brown) -- $\omega_k^{low-} = \omega_{k-k_c} - \omega_{k_c}$; line with short dashes (blue) -- linear dispersion relation $\omega_k=\sqrt{gk}$.}
\end{figure}

In the case of weak nonlinearity this rotation is the fastest process, but amplitude $A_{\vec k}(t)$
will be already slow function on time due to weakly nonlinear interaction of waves.
Let us investigate how the dispersion relation is influenced by the nonlinear
interactions in the system. For this we record $a_{\vec k}(t)$ for different values of $\vec k$ and then calculate a Fourier
transform on time, which leads to the function $a_{\vec k}(\omega)$. Because we have isotropic with respect to angle
situation, we can limit ourselves with harmonics on any ray starting at the origin of the $k$-plane.
In our numerical experiment we recorded
every other harmonics on the $k_x$-axis, from $\vec k=(4;0)$ till $\vec k=(28;0)$ and every tenth harmonic from $\vec k=(30;0)$ till $\vec k=(180;0)$ (for further wavenumbers the plot becomes extremely noisy, also it is already range of artificial dissipation) and calculated a Fourier transform. The resulting surface $|a_{\vec k}(\omega)|$ is represented in
Fig.~\ref{a_k_omega_surf}.
The observed sidebands are due to nonlinear interaction of waves with condensate. Although the nonlinear process
supposed to be weak, the fact, that condensate is more than an order of magnitude larger in amplitude (for harmonics further
from the condensate difference is even bigger) makes them essential. If we consider the simplest case of three-wave
nonlinear interaction of plane waves $a_{\vec k}(\vec r, t) = A_{\vec k}\exp[\I(\vec k\vec r -\omega t)]$, we shall get the following nonlinear terms:
\begin{align*}
a_{\vec k_0} a_{\vec k_c} = A_{\vec k_0} A_{\vec k_c} \exp \left[-\I(\omega(k_0)+\omega(k_c) +\I(\vec k_0 + \vec k_c)\vec r)\right],\\
a_{\vec k_0} a_{\vec k_c}^{*} = A_{\vec k_0} A_{\vec k_c}^{*} \exp \left[-\I(\omega(k_0)-\omega(k_c) +\I(\vec k_0 - \vec k_c)\vec r)\right],
\end{align*}
Here $\vec k_c$ - is the wavevector of condensate and $\vec k_0$ wavevector of some wave.
The first term corresponds to the upper sideband, second -- to the lower sideband.
For example, let us suppose that we measured frequency spectrum of some wave with wavevector $\vec k$, then it has to be
$\vec k = \vec k_0 + \vec k_c$ for the upper sideband and $\vec k = \vec k_0 - \vec k_c$ for the lower sideband.
Let us recall that we have isotropic situation, which means that our condensate in $k$-space is situated on a circle
with radius $k_c$ (really it is an annulus of finite width). Which means that for every case we have projection on the $k_x$ axis changing from $k_c$ to $-k_c$.
As a result our sidebands has to be between $\omega_k^{up-} = \omega_{k-k_c} + \omega_{k_c}$ and
$\omega_k^{up+} = \omega_{k+k_c} + \omega_{k_c}$ for the upper sideband and
between $\omega_k^{low-} = \omega_{k-k_c} - \omega_{k_c}$ and $\omega_k^{low+} = \omega_{k+k_c} + \omega_{k_c}$
for the lower sideband. Of course for the complete consideration we need to take into account the four wave interaction
terms. Most probably this is the reason for not so perfect agreement of our primitive estimate and the measurements
(Fig.~\ref{a_k_omega_surf}. On this stage of the investigation we are satisfied with this simple qualitative explanation.

\begin{figure}[ht!]
\centering
\includegraphics[width=3.4in]{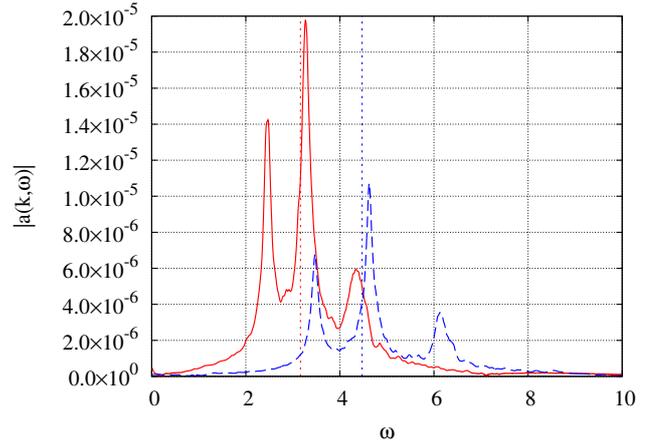}
\caption{\label{a_k_omega_010-020}Fig.~\ref{a_k_omega_010-020}. (Color online) Sections of the surface $|a_{\vec k}(\omega)|$. Solid line (red): spectral line for $\vec k = (10;0)$; dashed line (blue): spectral line for $\vec k = (20;0)$. Dotted vertical lines of both colors correspond to linear dispersion relation $\omega_k=\sqrt{gk}$.}
\end{figure}

If we consider spectrum in the vicinity of
condensate peak, one can see that sidebands are of close amplitudes with respect to central peak, corresponding to the
$\omega(k)=\sqrt{g k}$ dispersion relation (Fig.~\ref{a_k_omega_010-020}). Which means that at least for inverse cascade region distortion of the
dispersion relation have to be taken into account. For the direct cascade region, which is far from the condensate scale,
influence on the dispersion is almost negligible, although the central line is slightly shifted by the nonlinear frequency shift.

To summarize, we performed a direct numerical simulation of the isotropic turbulence of the surface gravity waves.
Currently, the inverse cascade slope, observed in numerical simulations,
is different from theoretical predictions of the wave turbulence theory. In order to investigate possible reasons
we measured the dispersion relation for waves in the presence of condensate. It was shown that in the region of inverse
cascade sidebands are of the same order of magnitude as the central line corresponding to the linear dispersion relation.
It means that in the vicinity of condensate we have to take the influence of the condensate into account. One of the possible
ways is to use the Bogolyubov transformation in order to calculate the augmented dispersion relation, in the same style
as it was done for phonons in liquid Helium. This is problem for future work, because in our case situation is
much more difficult, since condensate is located on the finite $k_c$ and coupling coefficient for gravity
waves is immensely more complex. We hope that our current result can be one of the building blocks for the theory of
inverse cascade in the presence of condensate.

The author would like to thank Yu.\,V.~Lvov and G.\,E.~Falkovich for enlightening discussions.
The idea about the Bogolyubov transformation was proposed by V.\,V.~Lebedev and I.\,V.~Kolokolov during
presentation of this work on the Scientific Council of the Landau Institute.
The author was partially supported by the NSF grant 1131791 and grant NSh-6885.2010.2.

Also the author would like to thank developers of FFTW~\cite{FFTW} and the whole GNU project~\cite{GNU} for developing, and supporting this useful and free software.


\end{document}